\documentclass[twocolumn,floatfix,showpacs,superscriptaddress]{revtex4}
\usepackage{graphicx}
\usepackage{amsmath}
\usepackage{bm}
\begin{document}

\title{Two-photon speckle as a probe of multi-dimensional entanglement}

\author{C. W. J. Beenakker}
\affiliation{Instituut-Lorentz, Leiden University, P.O. Box 9506, 2300 RA Leiden, The Netherlands}
\author{J. W. F. Venderbos}
\affiliation{Instituut-Lorentz, Leiden University, P.O. Box 9506, 2300 RA Leiden, The Netherlands}
\author{M. P. van Exter}
\affiliation{Huygens Laboratory, Leiden University, P.O. Box 9504, 2300 RA Leiden, The Netherlands}

\date{January 2009}
\begin{abstract}
We calculate the statistical distribution $P_{2}(I_{2})$ of the speckle pattern produced by a photon pair current $I_{2}$ transmitted through a random medium, and compare with the single-photon speckle distribution $P_{1}(I_{1})$. We show that the purity ${\rm Tr}\,\rho^{2}$ of a two-photon density matrix $\rho$ can be directly extracted from the first two moments of $P_{1}$ and $P_{2}$. A one-to-one relationship is derived between $P_{1}$ and $P_{2}$ if the photon pair is in an $M$-dimensional entangled pure state. For $M\gg 1$ the single-photon speckle disappears, while the two-photon speckle acquires an exponential distribution. The exponential distribution transforms into a Gaussian if the quantum entanglement is degraded to a classical correlation of $M\gg 1$ two-photon states. Two-photon speckle can therefore discriminate between multi-dimensional quantum and classical correlations.
\end{abstract}
\maketitle

Optical speckle is the random interference pattern that is observed when coherent radiation is passed through a diffusor or reflected from a rough surface. It has been much studied since the discovery of the laser, because the speckle pattern carries information both on the coherence properties of the radiation and on microscopic details of the scattering object \cite{Dai84,And05,Goo07}. The superposition of partial waves with randomly varying phase and amplitude produces a wide distribution $P(I)$ of intensities $I$ around the average $\langle I\rangle$. For full coherence and complete randomization the distribution has the exponential form $P(I)\propto\exp(-I/\langle I\rangle)$. The speckle contrast or visibility,
\begin{equation}
{\cal V}\equiv\langle I^{2}\rangle/\langle I\rangle^{2}-1,\label{Vdef}
\end{equation}
equals to unity for the exponential distribution.

These textbook results \cite{Man95} refer to \textit{single-photon} properties of the radiation, expressed by an observable $I_{1}$ that is quadratic in the field amplitudes. \textit{Biphoton} optics \cite{Khl88} is concerned with observables $I_{2}$ that are of fourth order in the field amplitudes, containing information on the entanglement of pairs of photons produced by a nonlinear optical medium. A variety of biphoton interferometers have been studied \cite{Jae93,Sal00,Kas03,Yar08}, but the statistical properties of the biphoton interference pattern produced by a random medium remain unknown. It is the purpose of this work to provide a theory for such ``two-photon speckle''.

There is a need for a such a theory, because of recent developments in the capabilities to produce entangled two-photon states of high dimensionality. The familiar \cite{Man95} polarization entangled two-photon state has dimensionality two and encodes a qubit \cite{Nie00}. Multi-dimensionally entangled two-photon states include spatial degrees of freedom \cite{Mai01,Wal03,Lan04,Nev05,Oem05,Ext07} and encode a ``qudit''. The dimensionality of the entanglement is quantified by the Schmidt rank $M$, which counts the number of pairwise correlated, orthogonal modes that have appreciable weight in the two-photon wave function \cite{Law04} and is an experimentally adjustable parameter \cite{Pee06}. 

As we will show in this paper, two-photon speckle not only provides information on the value of $M$, but it can also discriminate between quantum mechanical and classical correlations of $M$ modes. For classical correlations, on the one hand, the distributions $P_{1}(I_{1})$ and $P_{2}(I_{2})$ of single-photon and two-photon speckle both tend to narrow Gaussians upon increasing $M$ (with visibilities that vanish as $1/M$). For quantum correlations, on the other hand, $P_{1}$ tends to the same narrow Gaussian while $P_{2}$ becomes an exponential distribution.

\begin{figure}[tb]
\centerline{\includegraphics[width=0.8\linewidth]{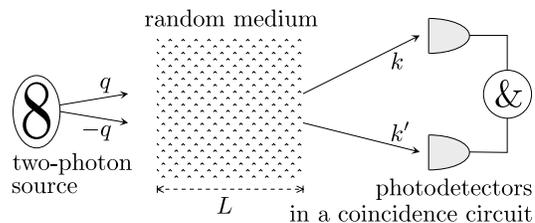}}
\caption{\label{fig_layout}
Schematic layout (not to scale) of a setup to detect two-photon speckle.
}
\end{figure}

We consider a monochromatic two-photon state of electromagnetic radiation (density operator $\hat{\rho}_{\rm in}$), scattered by a random medium (scattering matrix $S$). A pair of photodetectors in a coincidence circuit is located in the far field behind the random medium (see Fig.\ \ref{fig_layout}). The coincidence detection projects the scattered two-photon state (density operator $\hat{\rho}_{\rm out}$) onto quantum numbers $k$ and $k'$, which label the transverse wave vectors of an orthonormal basis of ${\cal N}$ modes. For a random medium of cross-sectional area ${\cal A}$ and for radiation of wave length $\lambda$, one has ${\cal N}\simeq\pi{\cal A}/\lambda^{2}$ per polarization degree of freedom. The spatial structure of the modes (and the precise value of ${\cal N}$) depends on the experimental geometry \cite{Law04,Pee06}, but all our statistical results are independent of it (for ${\cal N}\gg 1$) so we need not specify the modes further for our purpose.
 
In the far field (at a distance ${\cal D}\gg \sqrt{{\cal A}}$ from the random medium), the transmitted photon current $I_{1}(k)$ at a given $k$ is detected as a bright spot of area $\delta{\cal A}\simeq {\cal D}^{2}/{\cal N}\gg\lambda^{2}$. The random arrangement of bright and dark spots (the speckle pattern) depends sensitively on the realization of the randomness (for example, on the precise configuration of the scattering centra), and by varying the random medium \cite{note1} one samples a statistical distribution $P_{1}(I_{1})$.

The quantities $I_{1}(k)$ and $P_{1}(I_{1})$ refer to single-photon speckle. The biphoton current $I_{2}(k,k')$ counts the number of coincidence detection events per unit time, with one photon at $k$ and the other at $k'$. (We assume $k\neq k'$.) The distribution of $I_{2}$ in an ensemble of random realizations of the disorder is denoted by $P_{2}(I_{2})$ and describes two-photon speckle. Our goal is to find out what new information on the quantum state of the radiation can be extracted from $P_{2}$, over and above what is available from $P_{1}$.

The most general two-photon density operator at the input has the form
\begin{equation}
\hat{\rho}_{\rm in}=\tfrac{1}{2}\sum_{q_{1},q_{2}}\sum_{q'_{1},q'_{2}}\rho_{q_{1}q_{2},q'_{1}q'_{2}}a^{\dagger}_{q_{1}}a^{\dagger}_{q_{2}}|0\rangle\langle 0|a_{q'_{1}}a_{q'_{2}},\label{rhoindef}
\end{equation}
with $a^{\dagger}_{q}$ the photon creation operator in state $q$ and $|0\rangle$ the vacuum state. The coefficients in this expansion are collected in the ${\cal N}^{2}\times{\cal N}^{2}$ Hermitian density matrix $\rho$. Normalization requires that ${\rm Tr}\,\rho=1$. If the two-photon state is a pure state, then also ${\rm Tr}\,\rho^{2}=1$, while more generally the purity
\begin{equation}
{\cal P}={\rm Tr}\,\rho^{2}\in[0,1]\label{Pdef}
\end{equation}
quantifies how close the state is to a pure state \cite{Nie00}.

We will present an exact and general theory of the speckle statistics for arbitrary $\hat{\rho}_{\rm in}$, and also consider two specific simple examples: A maximally entangled pure state of Schmidt rank $M$,
\begin{equation}
\hat{\rho}_{\rm pure}=|\Psi_{M}\rangle\langle\Psi_{M}|,\;\;
|\Psi_{M}\rangle=M^{-1/2}\sum_{m=1}^{M}a^{\dagger}_{q_{m}}a^{\dagger}_{-q_{m}}|0\rangle,\label{rhopuredef}
\end{equation}
and its fully mixed counterpart
\begin{equation}
\hat{\rho}_{\rm mixed}=M^{-1}\sum_{m=1}^{M}a^{\dagger}_{q_{m}}a^{\dagger}_{-q_{m}}|0\rangle\langle 0|a_{q_{m}}a_{-q_{m}}.\label{rhomixeddef}
\end{equation}
Both states \eqref{rhopuredef} and \eqref{rhomixeddef} describe a pair of photons with anticorrelated transverse wave vectors \cite{note2}: If one photon has wave vector $q_{m}$, then the other photon has wave vector $-q_{m}$. (We assume $q_{m}\neq 0$ for each $m$.) The distinction between the two states is that the two photons in state \eqref{rhopuredef} are quantum mechanically entangled, while the correlation in state \eqref{rhomixeddef} is entirely classical. We will see how this difference shows up in the statistics of two-photon speckle.

Scattering by the random medium (in the absence of absorption) performs a unitary transformation on the creation and annihilation operators. If we collect the operators for the incident radiation in the vector $a$ and the operators for the scattered radiation in the vector $b$, then $b=S\cdot a\Leftrightarrow a=S^{\dagger}\cdot b$. Substitution into Eq.\ \eqref{rhoindef} gives the density operator of the outgoing state,
\begin{align}
\hat{\rho}_{\rm out}={}&\tfrac{1}{2}\sum_{q_{1},q_{2}}\sum_{q'_{1},q'_{2}}\rho_{q_{1}q_{2},q'_{1}q'_{2}}(S^{T}\cdot b^{\dagger})_{q_{1}}(S^{T}\cdot b^{\dagger})_{q_{2}}\nonumber\\
&\times|0\rangle\langle 0|(S^{\dagger}\cdot b)_{q'_{1}}(S^{\dagger}\cdot b)_{q'_{2}}.\label{rhooutdef}
\end{align}
From $\hat{\rho}_{\rm out}$ we obtain the biphoton current $I_{2}(k,k')$ by a projection,
\begin{equation}
I_{2}(k,k')=\tfrac{1}{2}\alpha_{2}{\rm Tr}\,\hat{\rho}_{\rm out}b^{\dagger}_{k}b^{\dagger}_{k'}b_{k}b_{k'},\label{I2def}
\end{equation}
where the coefficient $\alpha_{2}$ accounts for a nonideal detection efficiency and also contains the repetition rate of the photon pair production.

We now substitute Eq.\ \eqref{rhooutdef} into Eq.\ \eqref{I2def} to arrive at the required relation between the biphoton current and the scattering matrix,
\begin{equation}
I_{2}(k,k')=\alpha_{2}\sum_{q_{1},q_{2}}\sum_{q'_{1},q'_{2}}\rho_{q_{1}q_{2},q'_{1}q'_{2}}S_{kq_{1}}S_{k'q_{2}}S^{\ast}_{kq'_{1}}S^{\ast}_{k'q'_{2}}.\label{I2Srelation}
\end{equation}
Here we have assumed that $\rho$ is symmetric in both the first and second set of indices,
\begin{equation}
\rho_{q_{1}q_{2},q'_{1}q'_{2}}=\rho_{q_{2}q_{1},q'_{1}q'_{2}}=\rho_{q_{1}q_{2},q'_{2}q'_{1}}.\label{rhosymmetric}
\end{equation}
[We can assume this without loss of generality, since any antisymmetric contribution to $\rho$ would drop out of Eq.\ \eqref{rhoindef}.]

In order to compare with the single-photon current $I_{1}(k)$, we give the corresponding expressions,
\begin{equation}
I_{1}(k)=\tfrac{1}{2}\alpha_{1}{\rm Tr}\,\hat{\rho}_{\rm out}b^{\dagger}_{k}b_{k}=\alpha_{1}\sum_{q,q'}S_{kq}S^{\ast}_{kq'}\rho^{(1)}_{qq'},\label{I1def}
\end{equation}
in terms of the reduced single-photon density matrix
\begin{equation}
\rho^{(1)}_{qq'}=\sum_{q_{2}}\rho_{qq_{2},q'q_{2}}.\label{rho1def}
\end{equation}
The coefficient $\alpha_{1}$ is the single-photon detection efficiency (which may or may not be different from $\alpha_{2}$).

The next step is to calculate the statistical distributions $P_{1}$, $P_{2}$ of $I_{1}$, $I_{2}$. Following the framework of random-matrix theory \cite{Meh04,Vel04}, we make use of the fact that the matrix elements $S_{kq}$ have independent Gaussian distributions for ${\cal N}\gg 1$. (Corrections are of order $1/{\cal N}$.) The first moment vanishes, $\langle S_{kq}\rangle=0$, while the second moment depends on whether the radiation is detected in transmission or in reflection. In transmission through a random medium of length $L$ with mean free path $l$, one has
\begin{equation}
\langle |S_{kq}|^{2}\rangle=\frac{2l}{L{\cal N}}\equiv\sigma^{2}.\label{sigmadef}
\end{equation}

Let us begin by calculating the first two moments of $P_{1}$, $P_{2}$. Carrying out the Gaussian averages, we find for the mean values:
\begin{equation}
\langle I_{1}\rangle=\alpha_{1}\sigma^{2},\;\;
\langle I_{2}\rangle=\alpha_{2}\sigma^{4}\label{Iresult1}.
\end{equation}
(We omit the arguments $k$ and $k'$ for notational simplicity.) Neither mean value contains any information on the nature of the two-photon state. This is different for the variances ${\rm Var}\,I_{i}\equiv\langle I_{i}^{2}\rangle-\langle I_{i}\rangle^{2}$, for which we find
\begin{align}
&{\rm Var}\,I_{1}=\alpha_{1}^{2}\sigma^{4}\,{\rm Tr}\,\bigl(\rho^{(1)}\bigr)^{2},\label{I1result2}\\
&{\rm Var}\,I_{2}=\alpha_{2}^{2}\sigma^{8}\bigl[{\rm Tr}\,\rho^{2}+2\,{\rm Tr}\,\bigl(\rho^{(1)}\bigr)^{2}\bigr].\label{I2result2}
\end{align}
We conclude that the purity \eqref{Pdef} of the two-photon state can be obtained from the visibilities ${\cal V}_{i}\equiv ({\rm Var}\,I_{i})/\langle I_{i}\rangle^{2}$ of the single-photon and two-photon speckle patterns,
\begin{equation}
{\cal P}={\cal V}_{2}-2{\cal V}_{1}.\label{Presult}
\end{equation}
This is the first key result of our work.

To make contact with some of the literature on biphoton interferometry, we note that in the case of a pure two-photon state (when ${\cal P}=1$) knowledge of the single-photon visibility ${\cal V}_{1}$ fixes the two-photon visibility ${\cal V}_{2}$. The same holds (with some restrictions on the class of pure states and with a different definition of visibility) for the complementarity relations of Refs.\ \cite{Jae93,Sal00,Kas03,Yar08}. No such one-to-one relationship between ${\cal V}_{1}$ and ${\cal V}_{2}$ exists, however, for a mixed two-photon state.

We next turn to the full probability distribution $P_{2}$ of the two-photon speckle. Notice first that, if $\rho$ is far from a pure state, the ratio $\sqrt{{\cal V}_{2}}$ of the width of the distribution and the mean value is $\ll 1$. Indeed, for the fully mixed state \eqref{rhomixeddef} one has ${\rm Tr}\,\hat{\rho}_{\rm mixed}^{2}=1/M$ and ${\rm Tr}\,(\hat{\rho}_{\rm mixed}^{(1)})^{2}=1/2M$, so ${\cal V}_{2}=2/M\ll 1$ for $M\gg 1$. The relative magnitude of higher order cumulants is smaller by additional factors of $1/M$, hence $P_{2}$ tends to a narrow Gaussian for a fully mixed state with $M\gg 1$.

The situation is entirely different in the opposite limit of a pure state. The density matrix of a pure state factorizes,
\begin{equation}
\rho_{q_{1}q_{2},q'_{1}q'_{2}}=c_{q_{1}q_{2}}c^{\ast}_{q'_{1}q'_{2}},\label{rhopure}
\end{equation}
with $c$ a symmetric ${\cal N}\times {\cal N}$ matrix normalized by ${\rm Tr}\,cc^{\dagger}=1$. The corresponding reduced single-photon density matrix is $\rho^{(1)}=cc^{\dagger}$. The probability distributions $P_{2}$ and $P_{1}$ in this case of a pure two-photon state are related by an integral equation, which we derive in App.\ \ref{AppA}:
\begin{equation}
P_{2}(I_{2})=\Theta(I_{2})\frac{\alpha_{1}}{\alpha_{2}\sigma^{2}}\int_{0}^{\infty}dI_{1}\,\frac{P_{1}(I_{1})}{I_{1}}\exp\left(-\frac{\alpha_{1}}{\alpha_{2}\sigma^{2}}\frac{I_{2}}{I_{1}}\right).\label{P2P1relation}
\end{equation}
Here $\Theta(I)$ is the unit step function [$\Theta(I)=1$ if $I>0$, $\Theta(I)=0$ if $I<0$].

Without further calculation, we can conclude that when $P_{1}$ is narrowly peaked around the mean $\langle I_{1}\rangle$, the corresponding two-photon speckle distribution is the exponential distribution,
\begin{equation}
P_{2}(I_{2})\propto\exp\left(-\frac{\alpha_{1}}{\alpha_{2}\sigma^{2}}\frac{I_{2}}{\langle I_{1}\rangle}\right),\;\;{\rm if}\;\; {\cal V}_{1}\ll 1.\label{P2exp}
\end{equation}
The limiting exponential form is reached, for example, in the pure state \eqref{rhopuredef} for $M\gg 1$ (when ${\cal V}_{1}=1/2M\ll 1$). This is the second key result of our work.

\begin{figure}[tb]
\centerline{\includegraphics[width=0.9\linewidth]{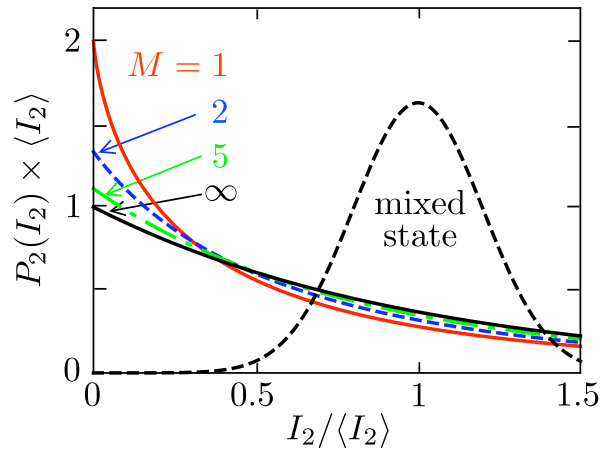}}
\caption{\label{fig_Kdistribution}
Probability distribution \eqref{P2Kresult} of the two-photon speckle for the maximally entangled pure state \eqref{rhopuredef} of Schmidt rank $M$. The exponential distribution \eqref{P2exp} (black solid curve) is reached in the limit $M\rightarrow\infty$. The black dashed curve shows the large-$M$ Gaussian distribution for the fully mixed state \eqref{rhomixeddef} (plotted for $M=50$).
}
\end{figure}

We can actually give a closed form expression for $P_{2}$ in terms of the eigenvalues of the matrix product $cc^{\dagger}$ (see App.\ \ref{AppB}), but it is rather lengthy. A more compact expression results for the special case of a maximally entangled pure state of Schmidt rank $M$ [Eq.\ \eqref{rhopuredef}]. Then all eigenvalues of $cc^{\dagger}$ are zero except a single $2M$-fold degenerate eigenvalue \cite{note3} equal to $1/2M$. The single-photon speckle distribution $P_{1}\propto I_{1}^{2M-1}\exp(-2MI_{1}/\alpha_{1}\sigma^{2})$ is a chi-square distribution with $4M$ degrees of freedom (since $I_{1}\propto\sum_{n=1}^{M}(|S_{k,q_{n}}|^{2}+|S_{k,-q_{n}}|^{2})$ is the sum of $2M$ Gaussian complex numbers squared). Substitution into Eq.\ \eqref{P2P1relation} leads to the following distribution of the two-photon speckle:
\begin{align}
P_{2}(I_{2})={}&\Theta(I_{2})\frac{4M}{\alpha_{2}\sigma^{4}(2M-1)!}\left(\frac{2MI_{2}}{\alpha_{2}\sigma^{4}}\right)^{M-1/2}\nonumber\\
&\times K_{2M-1}\left[2\sqrt{\frac{2MI_{2}}{\alpha_{2}\sigma^{4}}}\right].\label{P2Kresult}
\end{align}
The function $K_{2M-1}$ is a Bessel function. This distribution has appeared before in the context of wave propagation through random media \cite{And05} (where it is known as the ``$K$-distribution''), but there the parameter $M$ has a classical origin (set by the number of scattering centra) --- rather than the quantum mechanical origin which it has in the present context (being the Schmidt rank of the entangled two-photon state).

We have plotted the distribution \eqref{P2Kresult} for different values of $M$ in Fig.\ \ref{fig_Kdistribution}. The limiting value for $I_{2}\rightarrow 0$ equals
\begin{equation}
\lim_{I_{2}\rightarrow 0}P_{2}(I_{2})=\frac{2M}{(2M-1)\langle I_{2}\rangle}.\label{I2limit}
\end{equation}
The exponential form \eqref{P2exp} is reached quickly with increasing $M$ (black solid curve in Fig. \ref{fig_Kdistribution}). For comparison, we show in the same figure (black dashed curve) the Gaussian distribution reached for large $M$ in the case of the fully mixed two-photon state \eqref{rhomixeddef}. The striking difference with the entangled case is the third key result of our work.

In conclusion, we have presented a statistical description of the biphoton analogue of optical speckle. For an arbitrary pure state of two photons, the distribution $P_{2}$ of the two-photon speckle is related to the single-photon speckle distribution $P_{1}$ by an integral equation. A narrow Gaussian distribution $P_{1}$ maps onto a broad exponential distribution $P_{2}$. If the two-photon state is not pure, there is no one-to-one relationship between $P_{1}$ and $P_{2}$. For that case we show that knowledge of the visibilities of the single-photon and two-photon speckle patterns allows one to measure the purity of the two-photon state, thereby discriminating between classical and quantum correlations of $M$ degrees of freedom. All together, this theory provides a framework for the interpretation of ongoing experiments \cite{Pee09} on the propagation of multi-dimensionally entangled radiation through random media. 

We acknowledge discussions with W. H. Peeters and J. P. Woerdman. This research was supported by the Dutch Science Foundation NWO/FOM.

\appendix
\section{Derivation of the integral relation between $\bm P_{2}$ and $\bm P_{1}$ for a pure state}
\label{AppA}

We start from the expression \eqref{I2Srelation} for the biphoton current, which we write in the more compact form
\begin{equation}
I_{2}(k,k')=\alpha_{2}\,v'\cdot {\cal M}\cdot v'^{\ast},\label{I2Mrelation}
\end{equation}
in terms of a vector $v'$ with elements $v'_{q}=S_{k'q}$, and a matrix ${\cal M}$ with elements
\begin{equation}
{\cal M}_{qq'}=\sum_{q_{2},q'_{2}}v_{q_{2}}\rho_{qq_{2},q'q'_{2}}v^{\ast}_{q'_{2}}. \label{calMdef}
\end{equation}
(We have also defined a vector $v$ with elements $v_{q}=S_{kq}$.)

The probability distribution $P_{2}$ of $I_{2}$ is defined by
\begin{equation}
P_{2}(I_{2})=\left\langle\delta(I_{2}-\alpha_{2}\,v'\cdot {\cal M}\cdot v'^{\ast})\right\rangle_{v,v'},\label{P2delta}
\end{equation}
with Fourier transform
\begin{align}
F_{2}(\xi)&=\int_{-\infty}^{\infty}dI_{2}\,e^{i\xi I_{2}}P_{2}(I_{2})\nonumber\\
&=\langle\exp(i\xi\alpha_{2}\,v'\cdot {\cal M}\cdot v'^{\ast})\rangle_{v,v'}.\label{Fxidef}
\end{align}
The average $\langle\cdots\rangle_{v,v'}$ consists of an average over the scattering matrix elements $S_{kq}$ contained in the vector $v$ and an average over the scattering matrix elements $S_{k'q}$ contained in the vector $v'$. These two averages can be performed independently, since we have assumed $k\neq k'$. 

The Gaussian average over $v'$ can be carried out directly,
\begin{equation}
F_{2}(\xi)=\left\langle\frac{1}{{\rm Det}\,(1-i\xi\alpha_{2}\sigma^{2}{\cal M})}\right\rangle_{v}.\label{FxiDetM}
\end{equation}
To carry out the remaining average over $v$ we need to first evaluate the determinant. At this point we use that we are considering the density matrix $\rho$ of a pure state, which means that ${\cal M}$ factorizes as
\begin{equation}
{\cal M}_{qq'}=(c\cdot v)_{q}(c\cdot v)_{q'}^{\ast}.\label{calMpure}
\end{equation}
All eigenvalues of this matrix of rank $1$ vanish, except one nonzero eigenvalue $\mu_{1}$ given by
\begin{equation}
\mu_{1}=\sum_{q}|(c\cdot v)_{q}|^{2}=v\cdot\rho^{(1)}\cdot v^{\ast}=\frac{1}{\alpha_{1}}I_{1}(k),\label{mu1I1}
\end{equation} 
in view of the definition \eqref{I1def} of the single-photon current $I_{1}$.

Because of relation \eqref{mu1I1} the average over $v$ in Eq.\ \eqref{FxiDetM} may be replaced by an average over $I_{1}$,
\begin{equation}
F_{2}(\xi)=\int_{0}^{\infty}dI_{1}\,P(I_{1})\left(1-i\xi(\alpha_{2}/\alpha_{1})\sigma^{2}I_{1}\right)^{-1}.\label{FxiI1}
\end{equation}
Inverse Fourier transformation gives
\begin{align}
P(I_{2})&=\frac{1}{2\pi}\int_{-\infty}^{\infty}d\xi\,e^{-i\xi I_{2}}F_{2}(\xi)\nonumber\\
&=\Theta(I_{2})\frac{\alpha_{1}}{\alpha_{2}\sigma^{2}}\int_{0}^{\infty}dI_{1}\,\frac{P_{1}(I_{1})}{I_{1}}\exp\left(-\frac{\alpha_{1}}{\alpha_{2}\sigma^{2}}\frac{I_{2}}{I_{1}}\right).\label{P2P1relationApp}
\end{align}
This is the required relation \eqref{P2P1relation} between the distributions $P_{2}$ and $P_{1}$ of two-photon and single-photon speckle.
 
\section{Calculation of $\bm P_{2}$ for a pure state}
\label{AppB}

A general expression can be obtained for the distribution $P_{1}$ of single-photon speckle, by repeating the first few steps of App.\ \ref{AppA}:
\begin{align}
F_{1}(\xi)&=\int_{-\infty}^{\infty}dI_{1}\,e^{i\xi I_{1}}P_{1}(I_{1})\nonumber\\
&=\langle\exp(i\xi\alpha_{1}\,v\cdot \rho^{(1)}\cdot v^{\ast})\rangle_{v}\nonumber\\
&=\frac{1}{{\rm Det}\,(1-i\xi\alpha_{1}\sigma^{2}\rho^{(1)})}.\label{F1xia}
\end{align}
The reduced single-photon density matrix $\rho^{(1)}$ has distinct positive eigenvalues $\gamma_{m}$, each with multiplicity $\mu_{m}$ and satisfying the sum rule $\sum_{m}\mu_{m}\gamma_{m}=1$. In terms of these eigenvalues, Eq.\ \eqref{F1xia} can be written as
\begin{align}
F_{1}(\xi)&=\prod_{m}\left(1-i\xi\alpha_{1}\sigma^{2}\gamma_{m}\right)^{-\mu_{m}}\nonumber\\
&=\prod_{m}\frac{(-1)^{\mu_{m}-1}}{(\mu_{m}-1)!}\nonumber\\
&\qquad\times\left[\frac{d^{\mu_{m}-1}}{dx_{m}^{\mu_{m}-1}}(x_{m}-i\xi\alpha_{1}\sigma^{2}\gamma_{m})^{-1}\right]_{x_{m}\rightarrow 1}.\label{F1xib}
\end{align}
Inverse Fourier transformation gives the probability distribution,
\begin{widetext}
\begin{equation}
P_{1}(I_{1})=\Theta(I_{1})\sum_{m}\frac{1}{\alpha_{1}\sigma^{2}\gamma_{m}}\,\frac{(-1)^{\mu_{m}-1}}{(\mu_{m}-1)!}\,\left[\frac{d^{\mu_{m}-1}}{dx_{m}^{\mu_{m}-1}}\exp\left(-\frac{x_{m}I_{1}}{\alpha_{1}\sigma^{2}\gamma_{m}}\right)\prod_{m'\neq m}(1-x_{m}\gamma_{m'}/\gamma_{m})^{-\mu_{m'}}\right]_{x_{m}\rightarrow 1}.\label{P1gamman}
\end{equation}

This result for $P_{1}$ holds for any state of the radiation, but if we now assume that the state is pure, then we can use the relation \eqref{P2P1relation} to obtain the two-photon speckle distribution $P_{2}$ from $P_{1}$. (Note that, for a pure state, the $\gamma_{m}$'s are just the eigenvalues of the matrix product $cc^{\dagger}$.) Substitution of Eq.\ \eqref{P1gamman} into Eq.\ \eqref{P2P1relation} gives
\begin{equation}
P_{2}(I_{2})=\Theta(I_{2})\sum_{m}\frac{2}{\alpha_{2}\sigma^{4}\gamma_{m}}\frac{(-1)^{\mu_{m}-1}}{(\mu_{m}-1)!}\,\left[\frac{d^{\mu_{m}-1}}{dx_{m}^{\mu_{m}-1}}K_{0}\left(2\sqrt{\frac{x_{m}I_{2}}{\alpha_{2}\sigma^{4}\gamma_{m}}}\,\right)\prod_{m'\neq m}(1-x_{m}\gamma_{m'}/\gamma_{m})^{-\mu_{m'}}\right]_{x_{m}\rightarrow 1}.\label{P2gamman}
\end{equation}

The result \eqref{P2Kresult} given in the main text follows from Eq.\ \eqref{P2gamman} upon taking a single positive eigenvalue $\gamma_{1}=1/2M$ with multiplicity $\mu_{1}=2M$, so that
\begin{align}
P_{2}(I_{2})&=\Theta(I_{2})\frac{4M}{\alpha_{2}\sigma^{4}}\frac{(-1)^{2M-1}}{(2M-1)!}\,\left[\frac{d^{2M-1}}{dx^{2M-1}}K_{0}\left(2\sqrt{\frac{2MxI_{2}}{\alpha_{2}\sigma^{4}}}\,\right)\right]_{x\rightarrow 1}\nonumber\\
&=\Theta(I_{2})\frac{4M}{\alpha_{2}\sigma^{4}(2M-1)!}\left(\frac{2MI_{2}}{\alpha_{2}\sigma^{4}}\right)^{M-1/2}K_{2M-1}\left(2\sqrt{\frac{2MI_{2}}{\alpha_{2}\sigma^{4}}}\right).\label{P2pure}
\end{align}

\end{widetext}


\begin{thebibliography}{99}
\bibitem{note1}  In an experiment one would more conveniently sample the speckle distribution by taking different parts of a single speckle pattern, which for a sufficiently large number of speckles is equivalent to an average over configurations of scatterers.
\bibitem{note2} A state of the form \eqref{rhopuredef} would be produced, for example, by the nonlinear optical process of spontaneous type-I parametric down-conversion in the weak-focusing, noncollinear geometry. The two photons have the same frequency and the same linear polarization, while their transverse wave vectors are anticorrelated as $\pm q_{m}=\pm(q_{0}+\delta q_{m})$ (with $q_{0}\gg\delta q_{m}$). The Schmidt rank $M\simeq {\cal A}_{\rm beam}\Omega/\lambda^{2}$ is determined by the geometric extent  of the source, quantified by the near-field beam area ${\cal A}_{\rm beam}$ and the far-field opening angle $\Omega$.  See Ref.\ \protect\cite{Pee06}.
\bibitem{note3} The rank of $cc^{\dagger}$ is $2M$ rather than $M$ because $q_{m}$ and $-q_{m}$ contribute independently to the single-photon current (under the assumption that $q_{m}\neq 0$).
\end{thebibliography}
\end{document}